\documentclass{iau}
\usepackage{graphicx}

\title[Studying the age of supergiant companions in the magnetized X-ray binaries] 
{Studying the age of supergiant companions in the magnetized X-ray binaries}

\author[Ali Taani]   
{Ali Taani$^1$}

\affiliation{$^1$Physics Department, Faculty of Science, Al Balqa Applied University, 19117 Salt, Jordan \\ email: {\tt E-mail: ali.taani@bau.edu.jo} \\[\affilskip]}

\pubyear{2022}
\volume{xxx}  
\jname{Astronomy in Focus, Focus Meeting 4}
\editors{José Espinosa, ed.}

\begin{document}

\maketitle

\begin{abstract}
It has been proposed multiple times to use the neutron star (NS) in high-mass X-ray binaries (HMXBs) as an orbiting X-ray probe embedded in the wind-fed of its supergiant (SG) companion in order to constrain the stellar line-driven wind from the SG. We demonstrate how to combine various observables of HMXBs from the X-ray accretion luminosity produced by the wind-fed NS, in order to estimate and constrain the age of the donors. This would help us to study the stellar evolution track for each donor model. Since the evolution of massive stars is essentially determined by mass loss, and direct measures of mass-loss rates suffer from important uncertainties due to the unknown micro-structure of the wind.
\keywords{Binaries: X-rays: supergiant; wind-fed model; formation and evolution, magnetic fields.}
\end{abstract}

\firstsection 
\section{Introduction}

The cyclotron characteristics of electrons, protons, and other ions could be observed in a strong magnetic field, depending on the physical conditions in the emitting or absorbing matter (Coburn et al. 2002; Taani et al. 2019a,b). However, the question of where exactly the magnetic field is measured still remains unanswered. This depends on the accretion geometry and flow (Coburn et al. 2002; Mukherjee \& Bhattacharya 2012; Taani et al. 2020). As a result, their line profiles reflect the geometrical and physical properties during the accretion process on the neutron star (NS) surface. This work focuses on stellar-wind conditions in X-ray binary pulsars accompanied by supergiant companions.  From the masses and radii of the supergiant companions, which are derived from optical spectroscopies (Chaty et al. 2008; Reig et al. 2016; Mardini et al. 2020; Taani et al. 2022), wind velocities and mass-loss rates are estimated by using the empirical relations given by Hurley et al. (2000) and Vink et al. (2001). The obtained wind parameters are compared to those expected from X-ray observation results (luminosities and cyclotron-resonance spectral features) under the assumptions of the standard accretion spin-up model (e.g. Ghosh \& Lamb 1979; Cai et al. 2012; Mardini et al. 2022) and the spin-up/down equilibrium between accretion torque and propeller effect.

\section{Data Collection}
\begin{figure}[ht]
\begin{center}
 \includegraphics[width=0.4\textwidth]{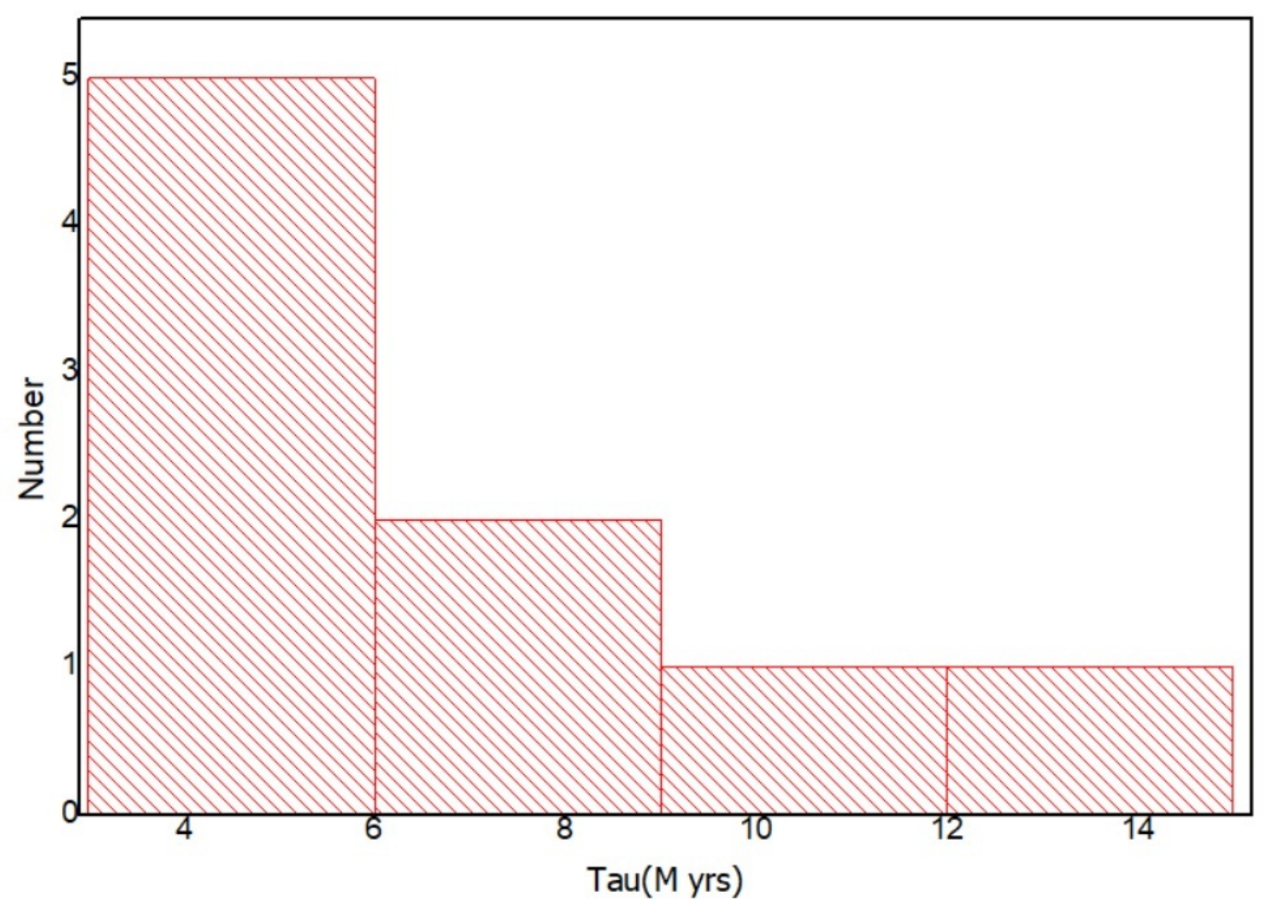}
 \caption{Histogram of the distribution of the age for the sample of SG analyzed in this work.}
 \label{fig1}
\end{center}
\end{figure}

The data on mass and radius come from the optical spectroscopy and the stellar- evolution model.
 We have taken the donor masses and radii of donors from Reig et al. (2016), Rawls et al. (2011) and Coley et al. (2015). The estimations depend on the empirical stellar-evolution model (Karino et al. 2009), and we computed the age of the donor. This would help us to study the stellar evolution track for each donor model. 
 The derived system ages are shown in Table~\ref{tab1}, where“$\tau$” in the unit of mega-year. 
 To illustrate all SG age distributions, a histogram of SG ages is plotted in Figure~\ref{fig1}. The distribution covers a span of
  (3-14) $\times$$10^{6}$ yrs.

\begin{table}
  \begin{center}
  \caption{The derived age for the supergiant companions.}
  \label{tab1}
 {\scriptsize
  \begin{tabular}{|l|c|c|c|c|c|}\hline 
name& $P_{spin}$ [s]& $P_{orb}$ [D]& $M_{donor}$($M_{\odot})$& $R_{donor}$($R_{\odot})$& $\tau$ [Myr]\\ \hline
    4U1538& 529&   3.73&  14.1&    12.5&   1 4\\
    Vela X-1&         283&  8.96& 24&  31.8&  7.3\\
    Cen X-3&         4.8&   2.09&  22.1&   12.6&  4.56\\
      LMC X-4  &        13.5&  1.4&  15&  7.7&  8.23\\
  4U0352 &        837&  250&  20&   13.2&  5.27\\
    4U2206&     5500&  19.1&  18&  8&  3.96\\
    2S0114&         9700& 11.6& 16& 37& 11.7\\
  J0440 &         205&  155&  17&  7.5&  4.26\\
    J16393&     604&   4.2&  20&   13&   5.42\\
    \hline
  \end{tabular}
  }
 \end{center}
\end{table}

\begin{figure}[ht]
\begin{center}
 \includegraphics[width=0.4\textwidth]{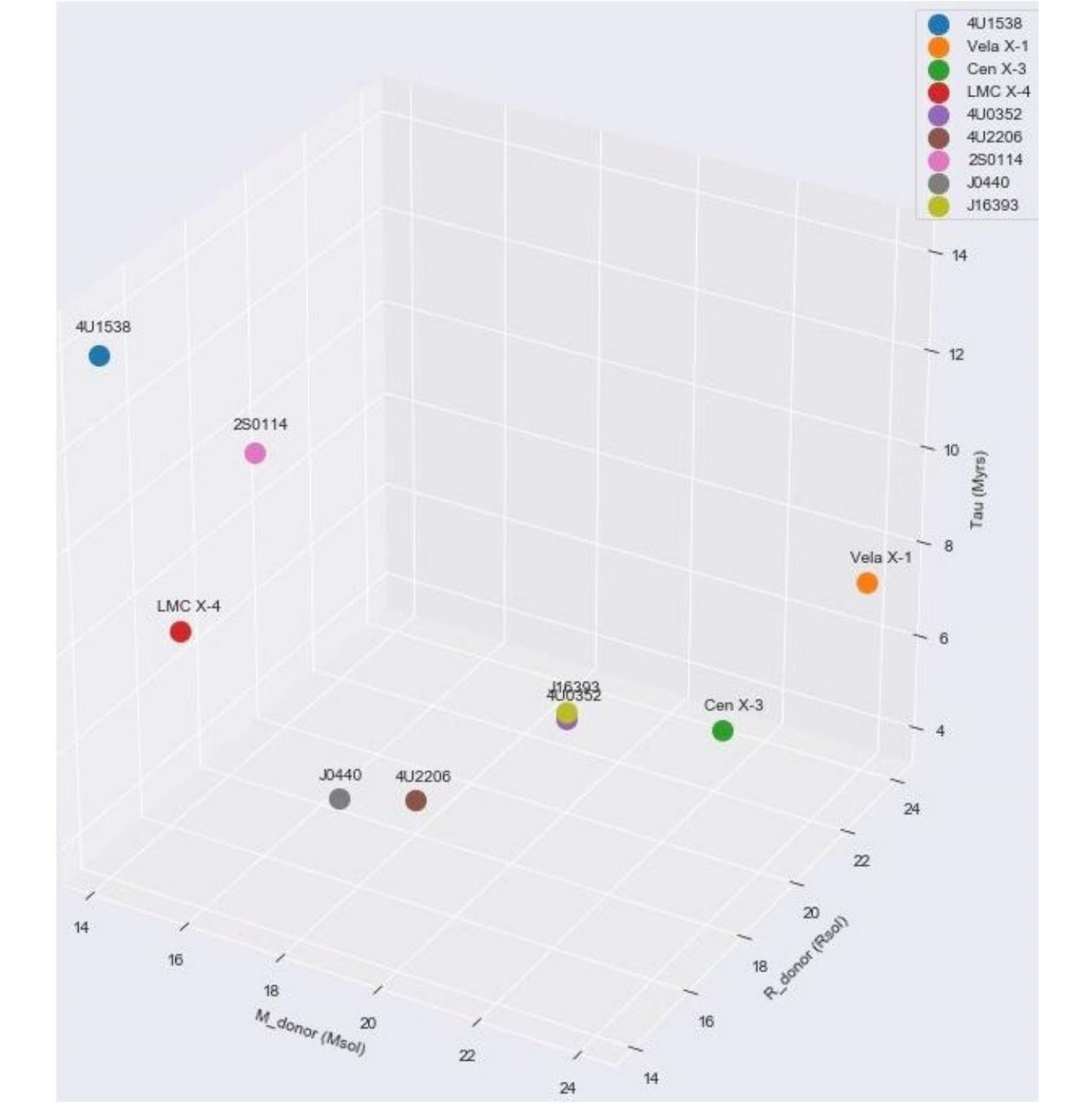}
 \caption{Illustrates the 3-D of the mass, radius and age for all the SGs.} 
 \label{fig2}
\end{center}
\end{figure}

It is worth noting that three sources (4U1907+09, GX301-2, and J16493-4348) have excessively large donors ($>$ 25~$M_{\odot}$). The stellar evolution pass, on the other hand, is almost entirely determined by their mass at zero-age. As a result, if the mass exchange episode can be neglected (it means that this scheme cannot be applied to massive donors. Figure~\ref{fig2}, illustrates the 3-D of the mass, radius and age for all the SGs. One should note that, it might be worthwhile  to check the relation between the magnetic field strength and the system age to see any interesting tendency. This result
should be confirmed by future observations.

\section{Conclusions}
We estimate the age of a number of donors in accreting systems using a set of assumptions about the accretion process, stellar wind properties, and other known observational properties of these sources (e.g., mass, radius, and luminosity). This would help us with the stellar evolution and computation of double compact object merger rates, as rightfully highlighted in this work. It is thus highly laudable to design semi-analytical models to do so, as is done in the present work. Several sources (4U1907+09, GX301-2, J16493-4348) have experienced mass exchanges in their binary evolution, and we cannot judge their mass at zero-age.
Hence, we cannot apply the stellar evolution code directly to these systems. This issue has to be confirmed and consolidated through additional optical spectroscopic observations.

\end{document}